\documentclass[prl,epsf,twocolumn]{revtex4}
\usepackage{graphicx}

\begin{document}

\title{Grain Boundary Scars and Spherical Crystallography}
\author{{\bf A. R. Bausch}$^{*}${\bf , M. J. Bowick}$^{\dag}${\bf , A.
Cacciuto}$^{\ddag}${\bf , A. D. Dinsmore}$^{\S}${\bf , M. F.
Hsu}$^{\P}${\bf , D. R. Nelson}$^{\P}${\bf , M. G.
Nikolaides}$^{*\P}${\bf , A.~ Travesset}$^{\parallel}$ \& {\bf D.
A. Weitz}$^{\P}$} \affiliation {$^{*}$ Department of Physics, E22,
Technische Universit{\"a}t M{\"u}nchen, 85747 M{\"u}nchen,
Germany} \affiliation{$^{\dag}$ Physics Department, Syracuse
University, Syracuse NY 13244-1130, USA } \affiliation{$^{\ddag}$
FOM Institute for Atomic and Molecular Physics, Kruislaan 407,
1098 SJ, Amsterdam, The Netherlands} \affiliation{$^{\S}$
Department of Physics, University of Massachusetts, Amherst, MA
01003-4525, USA} \affiliation{$^{\P}$ Department of Physics and
DEAS, Harvard University, Cambridge MA 02138, USA}
\affiliation{$^{\parallel}$ Physics and Astronomy Department, Iowa
State and Ames Natl. Lab, Ames, IA 50011, USA}

\begin{abstract}

{\bf We describe experimental investigations of the structure of
two-dimensional spherical crystals. The crystals, formed by beads
self-assembled on water droplets in oil, serve as model systems
for exploring very general theories about the minimum energy
configurations of particles with arbitrary repulsive interactions
on curved surfaces. Above a critical system size we find that
crystals develop distinctive high-angle grain boundaries, or
scars, not found in planar crystals. The number of excess defects
in a scar is shown to grow linearly with the dimensionless system
size. The observed slope is expected to be universal, independent
of the microscopic potential.}

\end{abstract}

\maketitle

Spherical particles on a flat surface pack most efficiently in a
simple lattice of triangles, similar to billiard balls at the
start of a game. Such six-fold coordinated triangular lattices
\cite{BM:1977} cannot, however, be wrapped on the curved surface
of a sphere; instead, there must be extra defects in coordination
number. Soccer balls and C$_{60}$ fullerenes
\cite{KHOCS:1985,J:2000} provide familiar realizations of this
fact {--} they have 12 pentagonal panels and 20 hexagonal panels.
The necessary packing defects can be characterized by their
topological or disclination charge, $q$, which is the departure of
their coordination number $c$ from the preferred flat space value
of 6 ($q=6-c$); a classic theorem of Euler
\cite{Euler:1750,HP:1996} shows that the total disclination charge
of any triangulation of the sphere must be 12 \cite{Sphere:12}. A
total disclination charge of 12 can be achieved in many ways,
however, which makes the determination of the minimum energy
configuration of repulsive particles, essential for
crystallography on a sphere, an extremely difficult problem. This
was recognized nearly 100 years ago by J.J. Thomson
\cite{JJT:1904}, who attempted, unsuccessfully, to explain the
periodic table in terms of rigid electron shells. Similar problems
recur in fields as diverse as multi-electron bubbles in superfluid
helium \cite{Leiderer:1995}, virus morphology
\cite{CasparKlug:1962,MarzecDay:1993,Viper:2001}, protein s-layers
\cite{Sleytr:2001,PMS:1991} and coding theory
\cite{Sloane:1984,ConwaySloane:1998}. Indeed, both the classic
Thomson problem, which deals with particles interacting through
the Coulomb potential, and its generalization to other interaction
potentials remain largely unsolved after almost 100 years
\cite{Smale:1998,Altschuler:1994/7,EH:1997}.

The spatial curvature encountered in curved geometries adds a
fundamentally new ingredient to crystallography, not found in the
study of order in spatially flat systems. To date, however,
studies of the Thomson and related problems have been limited to
theory and computer simulation. As the number of particles on the
sphere grows, isolated charge 1 defects are predicted to induce
too much strain; this can be relieved by introducing additional
dislocations, consisting of pairs of tightly bound 5-7 defects
\cite{Nelson:2002} which still satisfy Euler's theorem since their
net disclination charge is zero. Dislocations, which are
point-like topological defects in two dimensions, disrupt the
translational order of the crystalline phase but are less
disruptive of orientational order \cite{Nelson:2002}. While they
play an essential role in crystallography on a spherical surface,
the configuration and orientation of these excess defects remains
undetermined, and can only be fully understood through a
combination of theory and experiment. Experimental realizations
that probe the subtle structures have, however, been sorely
lacking.

We present an experimental realization of the generalized Thomson
problem that allows us to explore the lowest energy configuration
of the dense packing of repulsive particles on a spherical
surface.  We create two dimensional packings of colloidal
particles on the surface of spherical water droplets, and view the
structures with optical microscopy. Above a critical system size,
the thermally equilibrated colloidal crystals display distinctive
high-angle grain boundaries, which we label ``scars". These grain
boundaries are found to end within the crystal, which is not
observed to occur on flat surfaces because the energy penalty is
too high.

Our experimental system is based on the self-assembly of one
micron diameter cross-linked polystyrene beads adsorbed on the
surface of spherical water droplets (of radius $R$), themselves
suspended in a density-matched toluene/chlorobenzene mixture
\cite{DHNMBW:2002}. The particles are imaged with phase contrast
using an inverted microscope. The curvature of the spherical water
droplet limits the imaged surface area to between $5\%$ and $20\%$
of the full surface area of the sphere, depending on the size of
the droplet. After determining the center of mass of each bead,
the lattice geometry is analyzed by Delaunay triangulation
algorithms \cite{Delaunay:1934} appropriate to spherical surfaces.

\begin{figure}
\includegraphics[width=3in]{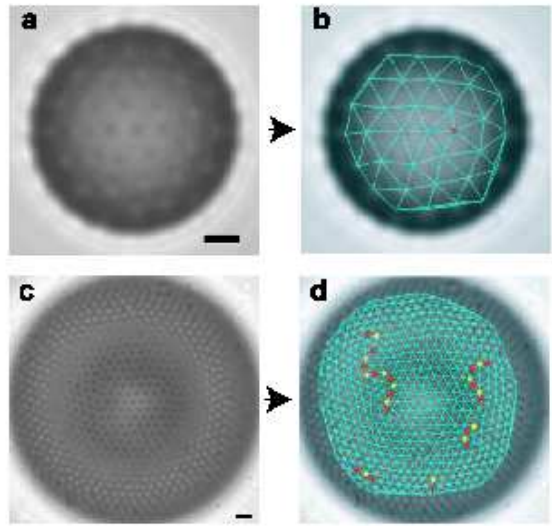}
\caption{Light microscope images of particle-coated droplets. Two
droplets ({\bf A}) and ({\bf C}) are shown, together with their
associated defect structures ({\bf B}) and ({\bf D}). Panel ({\bf
A}) shows an $\approx 13\%$ portion of a small spherical droplet
with radius $R = 12.0$ microns and mean particle spacing $a = 2.9$
microns (${\rm R/a} = 4.2$), along with the associated
triangulation ({\bf B}). Charge $+1(-1)$ disclinations are shown
in red and yellow respectively. Only one $+1$ disclination is
seen. Panel ({\bf C}) shows a cap of spherical colloidal crystal
on a water droplet of radius $R = 43.9$ microns with mean particle
spacing $a = 3.1 $ microns (${\rm R/a}=14.3$), along with the
associated triangulation ({\bf D}). In this case the imaged
crystal covers about $17\%$ of the surface area of the sphere. The
scale bars in ({\bf A}) and ({\bf C}) are 5 microns.}
\label{panel}
\end{figure}

We analyze the lattice configurations of a collection of 40
droplets.  A typical small spherical droplet with system size,
${\rm R/a}=4.2$, where $a$ is the mean particle spacing, is shown
in Fig.~\ref{panel}{\bf A}. The associated Delaunay triangulation
is shown in Fig.~\ref{panel}{\bf B}. The only defect is one
isolated charge $+1$ disclination. Extrapolation to the entire
surface of the sphere is statistically consistent with the
required $12$ total disclinations.

Qualitatively different results are observed for larger droplet
sizes as defect configurations with excess dislocations appear.
Although some of these excess dislocations are isolated, most
occur in the form of distinctive ($5-7-5-7-{\cdots}-5$) chains,
each of net charge $+1$, as shown in Fig.~\ref{panel}{\bf D}.
These chains form high-angle ($30^{\circ}$) grain boundaries, or
scars, which terminate freely within the crystal.  Such a feature
is energetically prohibitive in equilibrium crystals in flat
space. Thus, although grain boundaries are a common feature of 2D
and 3D crystalline materials, arising from a mismatch of
crystallographic orientations across a boundary, they usually
terminate at the boundary of the sample in flat space because of
the excessive strain energy associated with isolated terminal
disclinations. Termination within the crystal is a feature unique
to curved space. Thus, our results provide important guidance to
determine the configuration of excess defects on a sphere.

To ascertain that the colloidal particles are equilibrated, we use
particle tracking routines and subsequent automated triangulation
to measure the mobility of the observed screening dislocations.
The diffusion of thermally excited colloid particles on the
surface of the water droplets results in local rearrangements of
the crystal structure. Thermal fluctuations create and destroy
dislocations once every few seconds, on average, indicating that
the defect arrays reach equilibrium much faster than the
observation time of 10-60 min. The equilibration time can also be
estimated by the time required for a dislocation to diffuse across
typical defect structures. The measured diffusion constant allows
us to calculate equilibration times that range from a few seconds
to hundreds of seconds, depending on the size of the crystal.
Since the spherical crystals exist for 10 to 60 minutes, the
system has sufficient time to reach equilibrium. Thus, our
observations reflect the equilibrium ground state, as opposed to a
history-dependent nonequilibrium effect, which is the case for
crystals in flat space.

\begin{figure}
\includegraphics[width=2.5in]{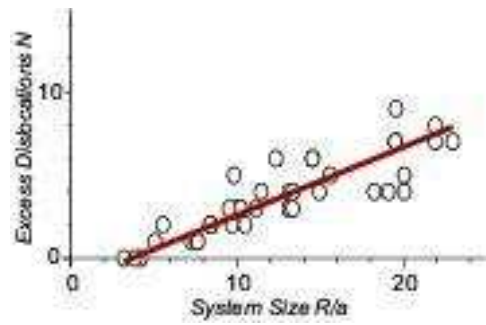}
\caption{Excess dislocations as a function of system size. The
number of excess dislocations per minimal disclination $N$ as a
function of system size $R/a$, with the linear prediction given by
theory shown as a solid red line.} \label{NvsR}
\end{figure}

To quantify the behavior of the scars, we determine the number of
excess dislocations per chain for each droplet, with the
convention that dislocations are counted as being part of the same
array if they are within three lattice spacings, and plot the
results as a function of ${\rm R/a}$ in Fig.~\ref{NvsR}. Scars
only appear for droplets with ${\rm R/a} \geq 5$. These results
provide a critical confirmation of a theoretical prediction that
${\rm R/a}$ must exceed a threshold value $({\rm R/a})_c \approx
5$, corresponding to $M \approx 360$ particles, for excess defects
to proliferate in the ground state of a spherical crystal
\cite{BNT:2000}. The precise value of $({\rm R/a})_c$ depends on
details of the microscopic potential but its origin is easily
understood by considering just one of the 12 charge $+1$
disclinations required by the topology of the sphere. In flat
space such a topological defect has an associated energy that
grows quadratically with the size of the system, since it is
created by excising a $60^{\circ}$ wedge of material and gluing
the boundaries together \cite{CL:1995}. The elastic strain energy
associated with this defect grows as the area. In the case of the
sphere the radius plays the role of the system size. As the radius
increases, isolated disclinations become much more energetically
costly. This elastic strain energy may be reduced by the formation
of linear dislocation arrays, i.e. grain boundaries. The energy
needed to create these additional dislocation arrays is
proportional to a dislocation core energy $E_c$ and scales
linearly with the system size. Such screening is inevitable in
flat space (the plane) if one forces an extra disclination into
the defect-free ground state. Unlike the situation in flat space,
grain boundaries on the sphere can freely terminate
\cite{BNT:2000,DM:1997,PGM:1999,BCNT:2002,Toomre}, and our
experimental results confirm these theoretical expectations.

\begin{figure}
\includegraphics[width=2.5in]{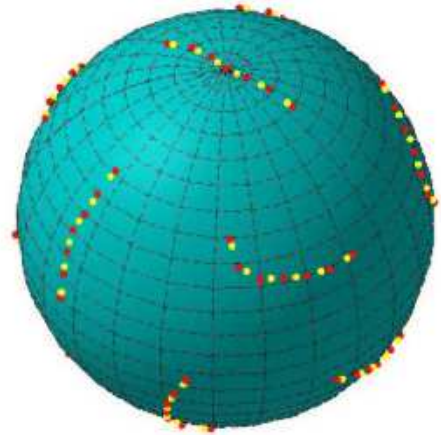}
\caption{Model grain boundaries. This image is obtained from a
numerical minimization, based on the theory of \cite{BNT:2000},
for a system size comparable to the droplet in
Figs.~\ref{panel}({\bf C},{\bf D}).} \label{theoryprediction}
\end{figure}

One systematic approach to determining the ground state of a
collection of $M$ particles distributed on the sphere and
interacting via an arbitrary repulsive potential
\cite{BNT:2000,BCNT:2002} treats the disclination defects
themselves as the fundamental degrees of freedom, with the
6-coordinated particles forming a continuum elastic background.
The agreement between the predicted and observed values of $({\rm
R/a})_c$ supports the validity of this theoretical approach. The
original particle pair potential is replaced by a long range
defect pair potential given by $\chi(\beta) \propto R^2(1 +
\int^{\frac{1-{\rm cos}\beta}{2}}_0 dz \frac{{\rm ln}z}{1-z})$,
for a pair of defects separated by an angular distance $\beta$.
The potential is attractive for opposite charged defects and
repulsive for like-charged defects. The underlying microscopic
potential enters only in determining the proportionality constant
(equivalent to an elastic Young modulus) and $E_c$. Many
predictions of this model are therefore universal in the sense
that they are insensitive to the exact microscopic potential. This
enables us to make definite predictions even though the colloidal
potential is not precisely known. It also means that our model
system serves as a prototype for any analogous system with
repulsive interactions and spherical geometry. To further test the
validity of this approach, we show a typical ground state for
large M in Fig.~\ref{theoryprediction}.  The system size here is
${\rm R/a}=12$, similar to the droplet in Fig.~\ref{panel}{\bf D}.
The results are remarkably similar to the experimentally observed
configuration in Fig.~\ref{panel}{\bf D}; the only difference is a
result of thermal fluctuations, which break the two defect scars
in the experiment. This agreement between theory and experiment
also provides convincing evidence that these scars are essential
components of the equilibrium crystal structure on a sphere.

The theory predicts that an isolated charge +1 disclination on a
sphere is screened by a string of dislocations of length ${\rm
cos}^{-1}(5/6)R \approx 0.59 R$ \cite{BNT:2000}. We can use the
variable linear density of dislocations to compute the total
number of excess dislocations $N$ in a scar. We find that $N$
grows for large $({\rm R/a})$ as $\frac{\pi}{3}\left[\sqrt{11} -
5\,{\rm cos}^{-1}(5/6)\right]\frac{\rm R}{\rm a} \approx
0.41(\frac{\rm R}{\rm a})$, independently of the microscopic
potential. This prediction is universal, and is in remarkable
agreement with the experiment, as shown by the solid line in
Fig.~\ref{NvsR}.

We expect these scars to be widespread in nature. They should
occur, and hence may be exploited, in sufficiently large viral
protein capsids, giant spherical fullerenes, spherical bacterial
surface layers (s-layers) and the siliceous skeletons of spherical
radiolaria (aulosphaera)\cite{ORA:1983}, provided that the
spherical geometry is not too distorted. Terminating strings of
heptagons and pentagons might serve as sites for chemical
reactions or even initiation points for bacterial cell division
\cite{Sleytr:2001} and will surely influence the mechanical
properties of spherical crystalline shells.

\noindent{\bf Acknowledgements}

\noindent This work was supported by the Department of Energy, the
National Science Foundation, the Harvard Materials Research
Science and Engineering Center (DMR-0213805), the Fonds der
Chemischen Industrie and the Emmy Noether Programme of the DFG.

\noindent Correspondence and requests for materials should be
addressed to MJB (email: bowick@physics.syr.edu) or ARB
(email:abausch@ph.tum.de).


\begin{thebibliography}{10}
\bibitem{Sphere:12} 12 here is obtained from $6\chi$, where the
Euler characteristic $\chi$ of a space is a topological invariant
equal to 2 for the sphere.
\end{thebibliography}
\end{document}